\documentclass[pre,twocolumn,aps,superscriptaddress]{revtex4}
\usepackage[T1]{fontenc}
\usepackage[latin9]{inputenc}
\usepackage{array}
\usepackage{booktabs}
\usepackage{braket}
\usepackage{mathrsfs}
\usepackage{mathbbol}
\usepackage{multirow}
\usepackage{amsmath}
\usepackage{amsthm}
\usepackage{amssymb}
\usepackage{stmaryrd}
\usepackage{graphicx}
\usepackage{latexsym}
\usepackage{textcomp}
\usepackage{mathtools}
\usepackage{xcolor}
\usepackage[citecolor=magenta,colorlinks=true]{hyperref}
\usepackage{lineno}
\usepackage{stackrel}
\usepackage[toc,page,titletoc]{appendix}
\usepackage{bm}
\definecolor{mycolor1}{rgb}{0.1, 0.6, 0.6}

\begin{document}


\title{
Generalized disorder averages and current fluctuations in run and tumble particles}
\author{Stephy Jose}
\email{stephyjose@tifrh.res.in}
\affiliation{Tata Institute of Fundamental Research, Hyderabad 500046, India}
\author{Alberto Rosso}
\email{alberto.rosso@universite-paris-saclay.fr}
\affiliation{LPTMS, CNRS, Univ.~Paris-Sud, Universit\'e Paris-Saclay, 91405 Orsay, France}
\author{Kabir Ramola}
\email{kramola@tifrh.res.in}
\affiliation{Tata Institute of Fundamental Research, Hyderabad 500046, India}

\date{\today}

\begin{abstract}
We present exact results for the fluctuations in the number of particles crossing the origin up to time $t$ in a collection of non-interacting run and tumble particles in one dimension. 
In contrast to passive systems, such active particles are endowed with two inherent degrees of freedom: positions and velocities, which can be used to construct density and magnetization fields. We introduce generalized disorder averages associated with both these fields and perform annealed and quenched averages over various initial conditions. 
We show that the variance $\sigma^2$ of the current in annealed versus quenched magnetization situations exhibits a surprising difference at short times: $\sigma^2 \sim t$ versus $\sigma^2 \sim t^2$ respectively, with a $\sqrt{t}$ behavior emerging at large times. Our analytical results demonstrate that in the strictly quenched scenario, where both the density and magnetization fields are initially frozen, the fluctuations in the current are strongly suppressed. Importantly, these anomalous fluctuations cannot be obtained solely by freezing the density field.
\end{abstract}

\maketitle

{\it Introduction:}  One-dimensional diffusive systems are known to exhibit a surprising characteristic in which they retain memory of their initial conditions indefinitely, i.e. even at late times the behavior of the system is influenced by how it was set up~\cite{le1989annealed,van1991fluctuations,derrida2009current2,ferrari2010interacting,lizana2010foundation,leibovich2013everlasting,krapivsky2012fluctuations,krapivsky2014large,banerjee2022role}. In analogy with disordered systems~\cite{mezard1987spin,binder1986spin,nishimori2001statistical,bouchaud1990anomalous}, there are two main types of initial conditions that are commonly studied. In the first, referred to as ``annealed'', averages are performed over all possible trajectories allowing for equilibrium fluctuations in the initial positions of particles while in the second, referred to as ``quenched'',  the initial positions of particles are held fixed~\cite{derrida2009current2,krapivsky2012fluctuations,cividini2017tagged,banerjee2022role,krapivsky2014large}. 
A quantity of central interest that is used to understand the effect of such initial conditions in one-dimensional stochastic systems is the integrated current i.e. the flux $Q$ of particles, across the origin up to time $t$. There have been several studies on the statistics of $Q$ for different systems such as a collection of non-interacting random walkers, symmetric simple exclusion process (SSEP), amongst others~\cite{derrida2007non,derrida2009current,derrida2009current2,krapivsky2012fluctuations,mallick2022exact,dandekar2022dynamical,dean2023effusion}. The mean of $Q$ which is a self-averaging quantity exhibits the same behavior for different initial conditions~\cite{krapivsky2012fluctuations}. However, the {\it fluctuations} of $Q$ strongly depend on how the system is set up at $t=0$, with annealed initial conditions displaying larger fluctuations~\cite{banerjee2022role}. Interestingly, the variance of $Q$ for these systems exhibits a $\sqrt{t}$ behavior at all times, with the coefficient determined by the type of initial conditions being employed. While there has been extensive research on current fluctuations in such passive systems~\cite{derrida2007non,derrida2009current,derrida2009current2,krapivsky2012fluctuations,mallick2022exact,derrida2004current,bodineau2006current,bodineau2007cumulants,dandekar2022dynamical,krapivsky2015tagged,dandekar2022macroscopic,dean2023effusion}, there have been relatively few studies of fluctuations in active systems~\cite{banerjee2020current,agranov2023macroscopic,jose2023current,di2023current}.

Active systems consist of particles that perform directed motion by consuming energy at the microscopic level, and represent an important paradigm in non-equilibrium physics~\cite{vicsek1995novel,czirok1999collective,tailleur2008statistical,cavagna2010scale,cates2012diffusive,ramaswamy2010mechanics}. Several microscopic models of active motion have been studied in detail in the literature \cite{evans2018run,malakar2018steady,mori2020universal,mori2020universalp,angelani2014first,martens2012probability,jose2022active,jose2022first,lindner2008diffusion,basu2018active,kumar2020active,romanczuk2010collective,romanczuk2012active,das2018confined,caprini2019active,sevilla2019stationary,le2019noncrossing}. A theoretically appealing class of active motion is the run and tumble particle (RTP) motion~\cite{evans2018run,malakar2018steady,mori2020universal,mori2020universalp,angelani2014first,martens2012probability,jose2022active,jose2022first}, in which an organism moves in a straight line (run) for a certain period of time, and then randomly changes direction (tumbles), before resuming another run. Recently~\cite{banerjee2020current}, it was shown that for non-interacting RTPs in one dimension, the variance of $Q$ displays a linear $t$ behavior at short times and a $\sqrt{t}$ behavior at large times. Similar to the case of passive particles, the coefficients governing these fluctuations differ for quenched and annealed initial conditions associated with the positions of the particles. However, in contrast to passive particles, whose motion is governed solely by random thermal fluctuations, the self-propulsion of active particles endows them with an {\it additional} degree of freedom: their internal bias direction. This opens up an intriguing possibility of constructing annealed and quenched disorder averages associated with both the positions and velocities of active particles.


In this Letter, we introduce generalized disorder averages for active particle systems which can lead to surprising differences in transport properties including the distribution of the particle flux $Q$ across the origin. Focusing on the case of non-interacting run and tumble particles in one dimension, we derive exact results for the fluctuations in $Q$ for each of the four types of disorder averages (quenched or annealed initial conditions for the positions and velocities respectively). 
Our analytic results demonstrate that for quenched initial conditions associated with {\it both} the positions and velocities of the particles, the current fluctuations are anomalously suppressed at short times. This suppression is characterized by a $t^2$ behavior in the variance of $Q$, as opposed to a $t$ behavior displayed in the other cases. This peculiar difference in growth exponents for different initial conditions does not seem to have an analogue in passive systems, where the initial conditions only give rise to different prefactors. 
Crucially, this anomalous suppression of fluctuations cannot be achieved by quenching the initial positions alone, highlighting the importance of considering both the positions and velocities of active particles in defining disorder averages.

{\it Microscopic model:}
We consider $N$ independent run and tumble particles evolving according to the Langevin equation 
\begin{equation}
\frac{\partial x_i(t)}{\partial t} = v m_i(t),~v>0,~1<i<N,
\label{langevin}
\end{equation}
in one dimension. 
Here, $m_i(t)$ is a stochastic variable that can switch values between $+1$ and $-1$ at a Poissonian rate $\gamma$. If $m_i(t)=+1$, the $i^{\text{th}}$ particle is in $+$ state at time $t$ and performs a biased motion towards the right. 
If $m_i(t)=-1$, the $i^{\text{th}}$ particle is in $-$ state at time $t$ and performs a biased motion towards the left. 
\begin{figure}[t!]
\includegraphics[width=0.9\linewidth]{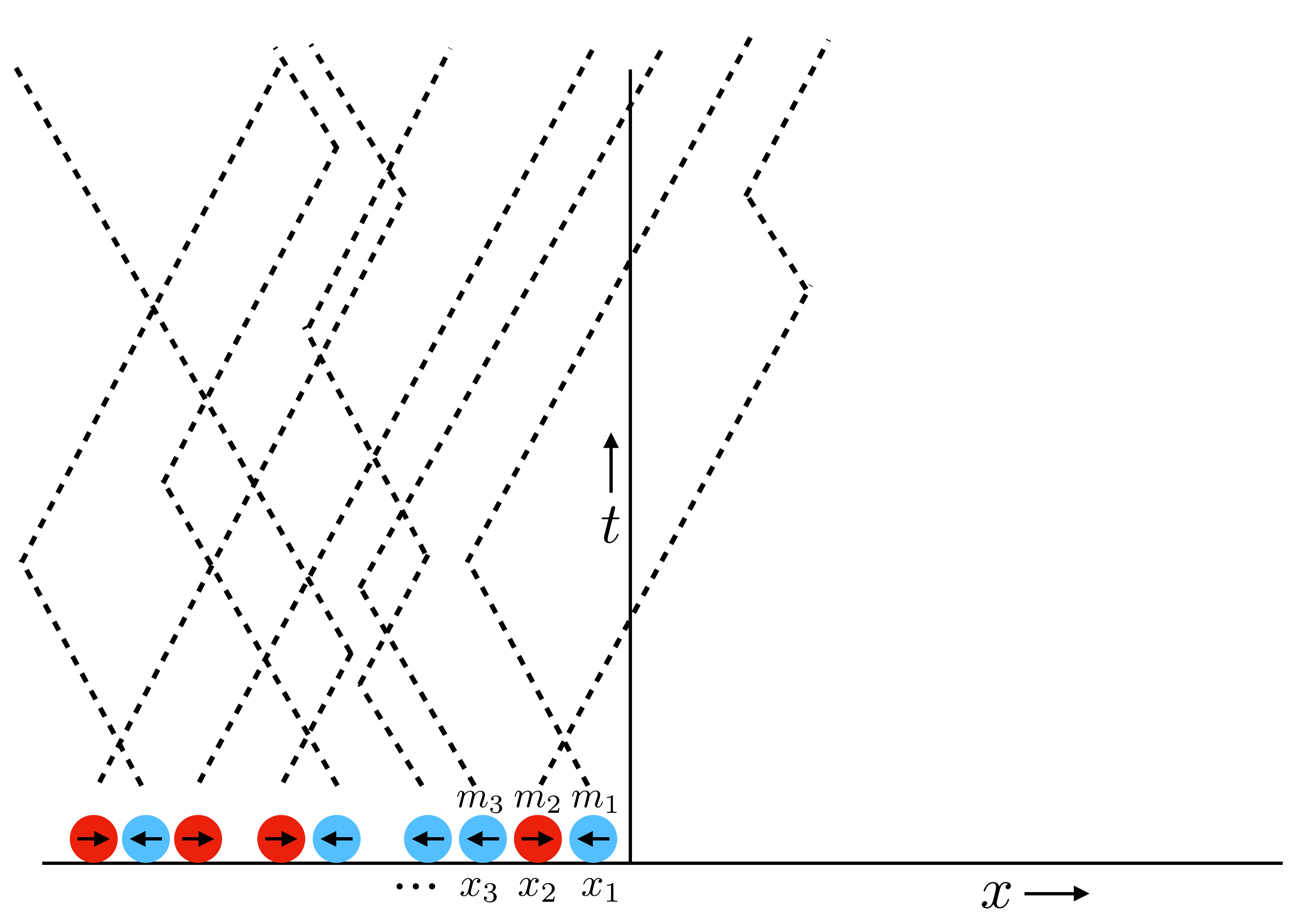}
\caption{Schematic diagram depicting the trajectories of non-interacting active particles in one dimension. The system is initiated using a step density profile, with all particles uniformly distributed to the left of the origin at time $t=0$. The initial positions and bias states of the particles are denoted by $\{x_i\}$ and $\{m_i \}$ respectively.}
\label{dynamics_fig}
\end{figure}
We consider a finite one-dimensional box bounded between $[-L,~0]$ with $N$ particles and then eventually take the infinite system size limit. The positions of the particles $x_i(t)$ can be used to construct a density field $\rho(x,t) = L^{-1}\sum_{i=1}^{N} \delta(x - x_i(t))$. Since the velocities resemble internal spin states, we construct a corresponding magnetization field $m(x,t) = L^{-1}\sum_{i=1}^{N} m_i(t) \delta(x - x_i(t))$ associated with the internal bias of the particles. For brevity, we denote the positions and bias states of particles at time $t=0$ by $\{x_i\}$ and $\{m_i \}$ respectively. Each position $x_i$ is drawn from a uniform distribution between $-L$ and $0$ with $0> x_1 > x_2 >x_3 \hdots$. The initial bias state $m_i$ can be $+$ or $-$ with probability $1/2$. This corresponds to a step initial density profile, $\rho(x,0) = \rho \theta(-x)$, where $\rho=N/L$ and a zero initial magnetization, $m(x,0) =0$. 

The number of particles $Q$, that cross the origin up to time $t$, can be computed as follows: when a particle crosses the origin from left to right, it contributes $+1$ to $Q$, and when a particle crosses from right to left, it contributes $-1$. Therefore, the integrated current up to time $t$ is exactly equal to the number of particles on the positive-half infinite line ($x>0$) at time $t$. The dynamics of a system of non-interacting RTPs is depicted in Fig.~\ref{dynamics_fig}.

{\it Summary of the main results:} 
We compute the statistics of $Q$ for various initial conditions. We first consider an annealed density and annealed magnetization setting. This allows for equilibrium fluctuations in the positions and velocities of particles at time $t=0$. This is the specific case studied in~\cite{banerjee2020current}. As we show, another equivalent scenario is the annealed density and quenched magnetization setting where the positions are allowed to fluctuate, but the velocities are fixed at time $t=0$.
The explicit expression for the variance of $Q$ for both these cases are the same and are given as~\cite{banerjee2020current}
\begin{eqnarray}
\sigma_{a,a}^2 = \sigma_{a,q}^2=   \frac{ \rho v}{2} t e^{-t \gamma } (\pmb{I}_0(t \gamma )+\pmb{I}_1(t \gamma )).
\label{var_a_density}
\end{eqnarray}
Here, $\pmb{I}_0$ and $\pmb{I}_1$ are modified Bessel functions.
The first and second subscripts denote the type of averaging done for the density and the magnetization fields respectively ($``a"$ for annealed and $``q"$ for quenched). 
The above expression has the limiting behaviors
\begin{equation}
\sigma_{a,a}^2 = \sigma_{a,q}^2 \approx
\begin{cases}
    \frac{ \rho v}{2}t,&\text{for } t \rightarrow 0,\\
     \rho \sqrt{\frac{{D_{\text{eff}}}~{t}}{{\pi}}},              &\text{for } t \rightarrow \infty,
\end{cases}
\end{equation}
where $D_{\text{eff}}=v^2/(2\gamma)$ is the effective diffusion constant for RTP motion in one dimension. The mean of $Q$, which is a self-averaging quantity assumes the same form for all initial conditions, and is also given by Eq.~\eqref{var_a_density}.
\begin{figure}[t!]
\includegraphics[width=0.9\linewidth]{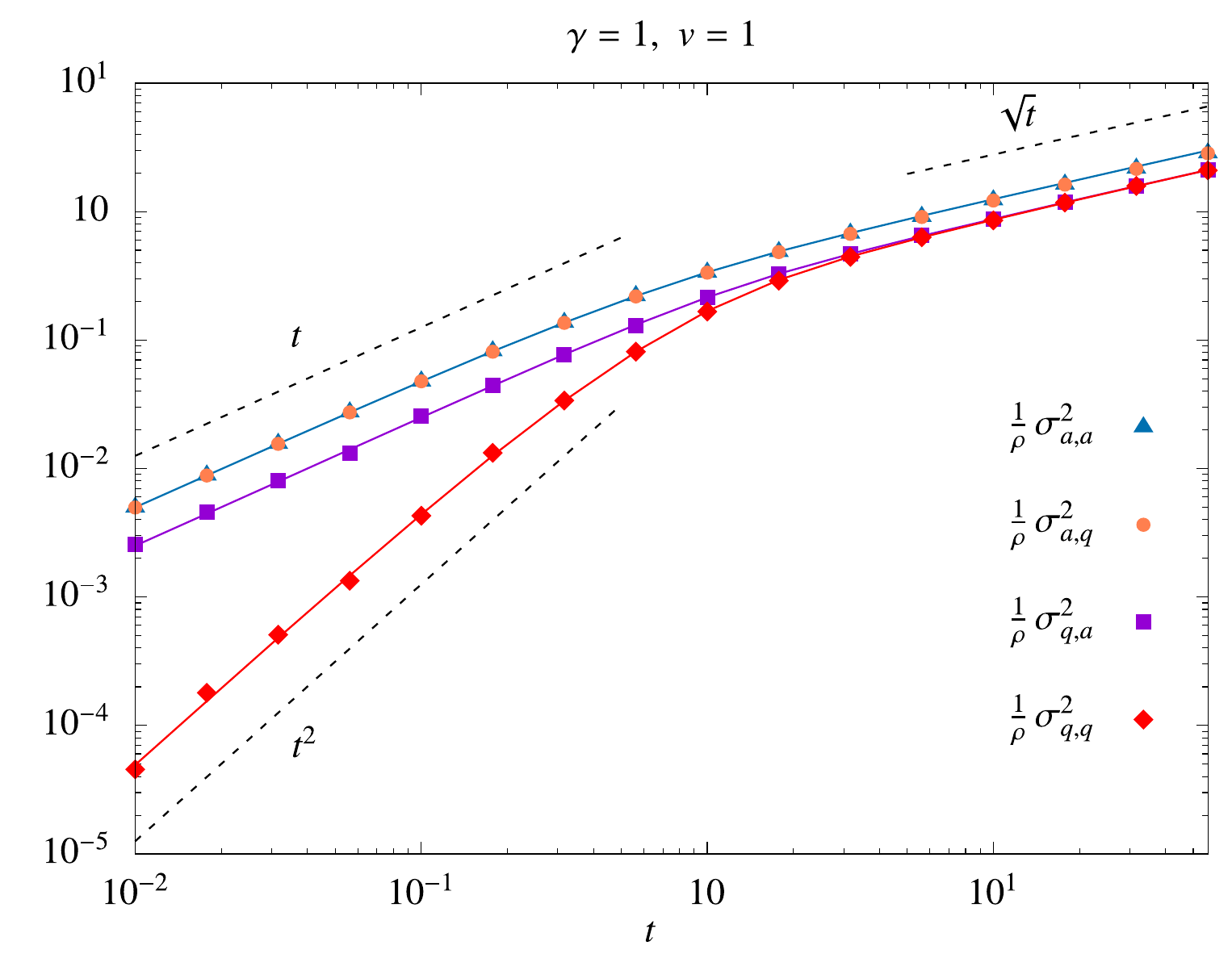}
\caption{Variance of the integrated current plotted as a function of time for different initial conditions. The solid curves correspond to the theoretical results provided in Eqs.~\eqref{var_a_density},~\eqref{var_q_q}~and~\eqref{var_q_a} and the points are from numerical simulations of the microscopic model. For quenched density and quenched magnetization initial conditions, the fluctuations surprisingly exhibit a $t^2$ behavior at short times.}
\label{var_a_q_fig}
\end{figure}

In the quenched density and quenched magnetization setting, the initial positions and velocities of the particles are held fixed. The variance $\sigma_{q,q}^2 $ for this case can be computed as

\begin{eqnarray}
 \sigma_{q,q}^2  &=& \frac{\rho v}{4} t  e^{-2 \gamma  t} \Big[(2+\pi  \pmb{L}_0(2 t \gamma )) \pmb{I}_1(2 t
 \gamma )\nonumber\\&-&\pi \pmb{L}_1(2 t \gamma ) \pmb{I}_0(2 t \gamma )\Big],
 \label{var_q_q}
\end{eqnarray}
where $\pmb{L}_0$ and $\pmb{L}_1$ are modified Struve functions. The above expression has the limiting behaviors
\begin{equation}
\sigma_{q,q}^2  \approx
\begin{cases}
    \frac{ \rho v \gamma}{2}t^2,&\text{for } t \rightarrow 0,\\
     \rho \sqrt{\frac{{D_{\text{eff}}}~{t}}{2{\pi}}},              &\text{for } t \rightarrow \infty.
\end{cases}
\end{equation}
We notice that for the case where both the fields are quenched initially, the fluctuations surprisingly exhibit a $t^2$ behavior at short times.

The final case we study is the quenched density and annealed magnetization setting where the initial positions of the particles are fixed, but the velocities are allowed to fluctuate. This non-trivial case is usually difficult to analyze, nevertheless we have exactly computed the variance $\sigma_{q,a}^2$ which has the explicit form
\begin{eqnarray}
 \sigma_{q,a}^2  &=&\frac{\rho v}{8} t  e^{-2 \gamma  t} \Big[(4+\pi  \pmb{L}_0(2 t \gamma )) \pmb{I}_1(2 t
 \gamma )\nonumber\\&+&(2-\pi  \pmb{L}_1(2 t \gamma )) \pmb{I}_0(2 t \gamma )\Big],
 \label{var_q_a}
\end{eqnarray}
with the limiting behaviors
\begin{equation}
\sigma_{q,a}^2  \approx
\begin{cases}
    \frac{ \rho v}{4}t,&\text{for } t \rightarrow 0,\\
     \rho \sqrt{\frac{{D_{\text{eff}}}~{t}}{2{\pi}}},              &\text{for } t \rightarrow \infty.
\end{cases}
\end{equation}
These asymptotic behaviors were also computed numerically in Ref.~\cite{banerjee2020current}. Our analytic expressions obtained in Eqs.~\eqref{var_a_density},~\eqref{var_q_q}~and~\eqref{var_q_a} are compared with direct numerical simulations in Fig.~\ref{var_a_q_fig}. We find excellent agreement between our theoretical predictions and the Monte Carlo simulation results.

In the calculations that follow, the angular bracket $\langle \cdots \rangle_{\{x_i\},\{m_i\}}$ denotes an average over the history (equivalent to a partition function), but with fixed initial positions $\{x_i\}$ and bias states $\{m_i\}$. We use $\overline{\cdots}$ to denote an average over initial positions and $\overbrace{\cdots}$ to denote an average over initial bias states. We consider different initial conditions separately.

{\it Annealed density and annealed magnetization:} 
We first consider annealed density and annealed magnetization initial conditions.  The flux distribution for this case is denoted as $P_{a,a}(Q,t)$. The moment-generating function for this distribution is given as
\begin{eqnarray}
&&\sum_{Q=0}^\infty e^{-p Q}  P_{a,a}(Q,t)  = \overbrace{ \overline{\langle e^{-p Q}\rangle_{\{x_i\},\{m_i\}}}} .
\end{eqnarray}
The quantity $\langle e^{-pQ}\rangle_{\{x_i\},\{m_i\}}$ appears in the expressions for the moment and cumulant generating functions of the flux distributions for different initial conditions. This quantity has been computed in~\cite{banerjee2020current} for the case where one field (density) is considered. We extend the definition to incorporate the magnetization field, which yields
\begin{equation}\label{history_av}
  \langle e^{-pQ}\rangle_{\{x_i\},\{m_i\}}= \prod_{i=1}^N\left[1- (1-e^{-p})U^{m_i}(-x_i,t)\right],
 \end{equation}
 where 
 \begin{eqnarray}
U^{m_i}(-x_i,t) = \int_0^\infty dx~ G^{m_i}(x,x_i,t),
\label{uzt_mi}
\end{eqnarray}
is the integral of the single-particle Green's function $G^{m_i}(x,x_i,t)$. The Green's function $G^{m_i}(x,x_i,t)$ gives the probability density of finding a particle at the location $x$ at time $t$, starting from the location $x_i$ in the bias state $m_i$ at time $t = 0$, and have not been derived previously. We present a detailed derivation of the Green's functions $G^{m_i}(x,x_i,t)$ for different initial bias states in the Supplemental Material~\cite{SI}.

To perform an average over initial positions, we consider the position of each particle to be distributed uniformly in the box $[-L,0]$, and eventually take a $L \rightarrow \infty,~N \rightarrow \infty$ limit with $N/L \rightarrow \rho$ fixed.
After performing an average over the initial positions in Eq.~\eqref{history_av}, we obtain
\begin{small}
\begin{eqnarray}
\overline{\langle e^{-p Q}\rangle_{\{x_i\},\{m_i\}}} &=& \prod_{i=1}^N\left[1- (1-e^{-p}) \overline{U^{m_i}(-x_i,t)}\right]\nonumber\\
&=& \left[1- \frac{1}{L}(1-e^{-p}) \int_0^L dz~U^{m_z}(z,t)  \right]^N .\nonumber\\
\label{history_pos_av}
\end{eqnarray}
\end{small}
Here, $m_z$ denotes the bias state of the particle located at $x_i=-z$ at time $t=0$.
Next performing an average over initial bias states in the above equation, we obtain
\begin{small}
\begin{eqnarray}
\overbrace{ \overline{\langle e^{-p Q}\rangle_{\{x_i\},\{m_i\}}}} &=& \left[1- \frac{1}{L}(1-e^{-p}) \int_0^L dz~U^{0}(z,t)  \right]^N,
\nonumber\\
\end{eqnarray}
\end{small}
where 
\begin{equation}
U^0(z,t)=\left( U^+(z,t)+U^-(z,t)\right)/2.
\label{U_0_def}
\end{equation}
Taking the limit $N \to \infty$, $L \to \infty$ keeping $\rho = N/L$ fixed yields
\begin{eqnarray}
\sum_{Q=0}^{\infty} e^{-pQ} P_{a,a}(Q,t) &=& \overbrace{ \overline{\langle e^{-p Q}\rangle_{\{x_i\},\{m_i\}}}} \nonumber\\&\rightarrow& \exp\left[-\mu(t) ~ (1-e^{-p})\right],
\end{eqnarray}
where
\begin{equation}
  \mu(t)=\rho \int_0^{\infty} dz ~  U^{0}(z,t).
  \label{mu_t}
\end{equation}
The above expression is exactly the moment-generating function for a Poisson distribution.
Therefore, $P_{a,a}(Q=N,t) $ is always a Poisson distribution~\cite{banerjee2020current} with
\begin{equation}
P_{a,a}(Q=N,t) = e^{-\mu(t)} \frac{\mu(t)^N}{N!},~ N~=0,1,2,\cdots~,
\end{equation}
and the mean and the variance are both given by $\mu(t)$. We thus obtain
\begin{equation}
    \sigma_{a,a}^2(t)=\mu(t).
    \label{equality_mu_t_sigma}
\end{equation}
The explicit expression for $\sigma_{a,a}^2$ can be directly computed using Eq.~\eqref{mu_t} and we obtain the announced result in Eq.~\eqref{var_a_density}. The derivation of this expression is provided in the Supplemental Material~\cite{SI}.

{\it Annealed density and quenched magnetization:} 
We next consider annealed density and quenched magnetization initial conditions. The flux distribution for this case is denoted as $P_{a,q}(Q,t)$. The moment-generating function for this flux distribution is given as
\begin{small}
\begin{eqnarray}
&&\sum_{Q=0}^\infty e^{-p Q}  P_{a,q}(Q,t)  = \exp \left[ \overbrace{\ln \overline{ \langle e^{-p Q}\rangle_{\{x_i\},\{m_i\}}}}\right] .
\end{eqnarray}
\end{small}
Using Eq.~\eqref{history_pos_av}, we directly compute the cumulant generating function as
\begin{small}
\begin{eqnarray}
&&\overbrace{\ln \overline{ \langle e^{-p Q}\rangle_{\{x_i\},\{m_i\}}}}=
\nonumber\\&&
\overbrace{ \ln \left[1- \frac{1}{L}(1-e^{-p}) \int_0^L dz~U^{m_z}(z,t)  \right]^N }\rightarrow-\mu(t)(1-e^{-p}),\nonumber\\
\end{eqnarray}
\end{small}
 where $\mu(t)$ is defined in Eq.~\eqref{mu_t}. In the large system size limit ($L \rightarrow \infty,~N \rightarrow \infty,~N/L \rightarrow \rho$), the distribution $P_{a,q}(Q,t)$ is equivalent to the distribution $P_{a,a}(Q,t)$.  This is because keeping the velocities fixed or allowing them to fluctuate does not make a difference in the annealed density setting as the initial positions of particles are randomized.  Thus we obtain the identity in Eq.~\eqref{var_a_density}.

{\it Quenched density and quenched magnetization:} 
The third case we study is the quenched density and quenched magnetization initial conditions. The flux distribution for this case is denoted as $P_{q,q}(Q,t)$. The moment-generating function is given as
\begin{small}
\begin{eqnarray}
&&\sum_{Q=0}^\infty e^{-p Q}  P_{q,q}(Q,t)  = \exp \left[\overbrace{ \overline{\ln \langle e^{-p Q}\rangle_{\{x_i\},\{m_i\}}}}\right] .
\end{eqnarray}
\end{small}
Taking the logarithm on both sides of Eq.~(\ref{history_av}) yields
\begin{small}
\begin{equation}
{\ln}\left[\langle e^{-pQ}\rangle_{\{x_i\},\{m_i\}} \right] = \sum_{i=1}^N {\ln}\left[1-(1-e^{-p})U^{m_i}(-x_i,t) \right].
\end{equation}
\end{small}
After performing an average over the initial positions and the velocities, we obtain the expression for the cumulant generating function as
\begin{small}
\begin{eqnarray}
\overbrace{\overline{\ln\left[\langle e^{-pQ}\rangle_{\{x_i\},\{m_i\}} \right]}}&=&  \frac{\rho}{2}  \int_0^\infty dz \ln \left[ 1 - (1-e^{-p}) U^+(z,t)\right] \nonumber\\&+&\frac{\rho}{2} \int_0^\infty dz \ln \left[ 1 - (1-e^{-p}) U^-(z,t)\right] .
\nonumber\\
\end{eqnarray} 
\end{small}
The cumulants can be extracted by collecting terms that appear in the same powers of $p$. This yields the expressions for the mean and variance of $Q$ as
\begin{eqnarray}
\langle Q \rangle_{q,q}  &=&\mu(t), \nonumber \\
\sigma_{q,q}^2 &=& \langle Q^2 \rangle_{q,q} - \langle Q \rangle^2_{q,q} \nonumber\\
&=& \frac{\rho}{2} \int_0^\infty dz ~\left[ U^+(z,t)(1-U^+(z,t)) \right] \nonumber\\&+& \frac{\rho}{2} \int_0^\infty dz~\left[ U^-(z,t)(1-U^-(z,t)) \right]  ,
\label{var_q_q_def}
\end{eqnarray}
where $\mu(t)$ is defined in Eq.~\eqref{mu_t} and the explicit expression for $\mu(t)$ is provided in Eq.~\eqref{var_a_density}. One can also derive the explicit expression for the variance using Eq.~\eqref{var_q_q_def}. Thus we obtain the announced result in Eq.~\eqref{var_q_q}. See \cite{SI} for details related to the derivation of this expression.

{\it Quenched density and annealed magnetization:} 
Finally, we study quenched density and annealed magnetization initial conditions. The flux distribution for this case is denoted as $P_{q,a}(Q,t)$. The moment-generating function for this process is given as
\begin{small}
\begin{eqnarray}
&&\sum_{Q=0}^\infty e^{-p Q}  P_{q,a}(Q,t)  = \exp \left[ \overline{\ln \overbrace{ \langle e^{-p Q}\rangle_{\{x_i\},\{m_i\}}}}\right] .
\end{eqnarray}
\end{small}
After performing an average over the initial velocities in Eq.~\eqref{history_av}, we obtain
\begin{eqnarray}
\overbrace{\langle e^{-p Q}\rangle_{\{x_i\},\{m_i\}}} &=& \prod_{i=1}^N\left[1- (1-e^{-p}) \overbrace{U^{m_i}(-x_i,t)}\right]\nonumber\\&=&\prod_{i=1}^N\left[1- (1-e^{-p}) U^0(-x_i,t)\right],\nonumber\\
\end{eqnarray}
where $U^0$ is defined in Eq.~\eqref{U_0_def}. Next, we compute the cumulant generating function as
\begin{small}
\begin{eqnarray}
&&\overline{\ln \overbrace{ \langle e^{-p Q}\rangle_{\{x_i\},\{m_i\}}}}\nonumber\\&&= \frac{N}{L}\int_{-L}^0 dx_i ~{\ln}\left[1-(1-e^{-p})U^0(-x_i,t) \right] \nonumber\\ 
&&\longrightarrow \rho \int_0^\infty dz \ln \left[ 1 - (1-e^{-p}) U^0(z,t)\right].\nonumber
\end{eqnarray}
\end{small}
Collecting terms that appear in the first and second powers of $p$, we obtain the expressions for the mean and variance of $Q$ as
\begin{eqnarray}
\langle Q \rangle_{q,a} &=&\mu(t),  
\nonumber\\
\sigma_{q,a}^2 &=& \langle Q^2 \rangle_{q,a} - \langle Q \rangle^2_{q,a} \nonumber\\
&=& \rho \int_0^\infty dz~\left[ U^0(z,t)\left(1-U^0(z,t)\right)\right]  , 
\label{var_q_a_def}
\end{eqnarray}
where $\mu(t)$ is defined in Eq.~\eqref{mu_t} and the explicit expression for $\mu(t)$ is provided in Eq.~\eqref{var_a_density}. The explicit expression for the variance can also be computed using the single-particle Green's functions~\cite{SI} and we obtain the announced result in Eq.~\eqref{var_q_a}.

{\it Discussion}:
In this Letter, we have introduced generalized disorder averages for active systems that account for fluctuations in both the initial positions and velocities of particles. We illustrated these averages in a one-dimensional system of non-interacting RTPs, and derived exact results for the fluctuations in the integrated current across the origin up to time $t$. Surprisingly, these fluctuations display different growth exponents at short times for different initial conditions, a feature that does not seem to occur in passive systems. Specifically, we observed suppressed fluctuations for the case when both the positions and velocities are initially quenched, characterized by a $t^2$ growth in the variance, as opposed to a linear $t$ behavior observed in the other cases. At large times, a $\sqrt{t}$ behavior emerges, consistent with late time diffusive behavior, with the quenched and annealed density settings differing by a factor of $\sqrt{2}$~\cite{leibovich2013everlasting,banerjee2020current,derrida2009current2,krapivsky2012fluctuations}.

Such generalized disordered averages can also be extended to a variety of other models with multiple degrees of freedom at the particle level. While the exact results reported in this Letter are specific to non-interacting RTPs in one dimension, we expect the derived asymptotic behaviors to hold in other active systems as well. In particular, we expect other models of active motion such as active Brownian particles~\cite{lindner2008diffusion,basu2018active,kumar2020active,romanczuk2010collective,romanczuk2012active} as well as interacting active particles in the low-density limit~\cite{kourbane2018exact,agranov2021exact,jose2023current,agranov2022entropy} to display similarly suppressed fluctuations for quenched initial conditions. Additionally, we also expect the surprising differences in the growth exponents governing the current fluctuations for different initial conditions to appear in higher dimensions as well. 

Several interesting directions for further research remain. It would be instructive to compute the higher-order cumulants of the current, in order to better understand the differences between quenched and annealed disorder in active systems. It would also be interesting to verify our predictions by implementing the different disorder averages within a fluctuating hydrodynamics framework or using Macroscopic Fluctuation Theory~\cite{bertini2005current,bertini2006non,bertini2007stochastic,bertini2009towards,bertini2015macroscopic,derrida2001free,bodineau2004current,derrida2011microscopic}, which have been shown to successfully predict fluctuations in many-particle systems~\cite{derrida2009current2,krapivsky2012fluctuations,krapivsky2015tagged,krapivsky2014large,mallick2022exact,derrida2007non,rana2023large}. Additionally, although we have considered the case of zero diffusion and symmetric initial conditions, our framework can also be extended to situations where the particles have a non-zero diffusion in their microscopic dynamics, and are initiated with asymmetric magnetization initial conditions. Finally, it would also be interesting to study the effect of generalized disorder averages on observables such as the magnetization current, which could lead to a better understanding of the differences between active and passive systems.

{\it Acknowledgments:} We thank R. Maharana, M. Barma, R. Dandekar, S. N. Majumdar and G. Schehr for useful discussions. The work of K.~R. was partially supported by the SERB-MATRICS grant MTR/2022/000966. This project was funded by intramural funds at TIFR Hyderabad from the Department of Atomic Energy (DAE), Government of India.


\bibliography{Current_fluctuations}

\clearpage

\begin{widetext}

\section*{\large Supplemental Material for\\ ``Generalized disorder averages and current fluctuations in run and tumble particles''}

In this document, we provide supplemental details related to the results presented in the main text. We provide expressions for single particle propagators for different initial conditions. We consider symmetric initial conditions where a Run and Tumble Particle (RTP) is initialized in $+$ and $-$ states with equal probability as well as asymmetric initial conditions where a RTP is initialized in either the $+$ or $-$ states. We also present detailed derivations of the exact expressions for the variance of the integrated current for quenched and annealed initial conditions involving both the density and magnetization fields.

\maketitle


\section{Single Particle Propagators}
Consider an RTP starting its motion from the location $x=0$ at time $t=0$ in one dimension. Let  $P_m(x,t)$ denote the probability density that the particle is at the location $x$ at time $t$ in the  velocity state $m$, where $m=\pm $. The evolutions equations for the probability density read~\cite{malakar2018steady}
\begin{eqnarray}
\frac{\partial P_+(x,t)}{\partial t}
&=& -v \frac{\partial P_+(x,t)}{\partial x} - \gamma P_+(x,t) + \gamma P_-(x,t),  \nonumber\\
\frac{\partial P_-(x,t)}{\partial t}
&=& +v \frac{\partial P_-(x,t)}{\partial x} - \gamma P_-(x,t) + \gamma P_+(x,t).  \label{Pequations}
\end{eqnarray}
The total probability density to be at location $x$ at time $t$ is defined as 
\begin{equation}
    P(x,t)=P_+(x,t)+P_-(x,t).
\end{equation}
Let us define the Fourier transform of the probability density $P_{m}(x,t)$ as
\begin{equation}
\tilde P_m(k, t) = \int_{-\infty}^\infty~d k { e}^{i k x} P_m (x,t),
\end{equation}
and the inverse Fourier transform as
\begin{equation}
P_m (x,t) = \frac{1}{2 \pi}\int_{-\infty}^\infty~d x { e}^{-i k x} \tilde P_m(k, t).
\end{equation}
Taking a Fourier transform of Eq.~\eqref{Pequations} yields the matrix equation
\begin{equation}
\label{matrix11}
\frac{\partial}{\partial t}
\ket{\tilde P_m (k,t)}
=
\mathcal{A}(k)\ket{\tilde P_m (k,t)},
\end{equation}
where the ket $\ket{\tilde P_m (k,t)}$ denotes the column vector ${\begin{pmatrix}
\tilde P_+(k,t)\\
\tilde P_-(k,t)
\end{pmatrix}}$,
and the matrix $\mathcal{A}(k)$ is given as 
\begin{equation}
\label{matrix2}
\mathcal{A}(k)=
\begin{pmatrix}
 -\gamma+ivk & \gamma \\
 \gamma & -\gamma-ivk 
 \end{pmatrix}.
\end{equation}
We next define the Laplace transform of $P_{m}(x,t)$ as
\begin{equation}
\tilde P_m(x, s) = \int_0^\infty~d t { e}^{-st} P_m (x,t).
\end{equation}
Taking a Laplace transform of Eq.~(\ref{matrix11}) yields
\begin{equation}
\label{matrix11a}
\left [s\mathcal{I}-\mathcal{A}(k)\right]~
\ket{\tilde P_m (k,s)}
= \ket{\tilde P_m (k,t=0)},
\end{equation}
where $\mathcal{I}$ is the two dimensional identity matrix.  
\subsection{Symmetric initial conditions}
We first consider the case of a single RTP initiated symmetrically in the states, $+$ and $-$. At time $t=0$, the particle has an equal probability to be in the state $+$ or $-$. Therefore,
\begin{equation}
P_+(x,0) = P_-(x,0) = \frac{1}{2} \delta(x).
\label{sym_cond}
\end{equation}
In Fourier space, these initial conditions translate to $\ket{\tilde P_m (k,t=0)}~=~{\begin{pmatrix}
1/2~
1/2
\end{pmatrix}}^T$.
Solving Eq.~(\ref{matrix11a}) along with these initial conditions yields the exact expression for the Fourier-Laplace transform of the total probability density as
\begin{equation}
\label{fl}
\tilde P (k,s)=\frac{s+2 \gamma }{s (s+2 \gamma )+v^2 k^2},
\end{equation}
where  $\tilde P(k,s)~=~\tilde P_+(k,s)+\tilde P_-(k,s)$.
We next invert the Fourier transform in the above equation to obtain
\begin{equation}
\tilde P(x,s) =  {\rm e}^{-|x|\lambda}\frac{\lambda }{2s},
\label{ptot_sym}
\end{equation}
where $\lambda$ is defined as
\begin{equation}
\lambda =  \frac{\sqrt{s(s+2\gamma)}}{v}.
\label{lambda}
\end{equation}

Let us define the Green's function $G^0(x,x_i,t)$ as the probability density of finding an RTP at location $x$ at time $t$, starting from the location $x_i$ at time $t = 0$ with equal probabilities to be in $+$ or $-$ state. Here, the superscript $ ``0"$ denotes the symmetric case where the particle has an equal probability to in $+$ or $-$ state at time $t=0$. Since the evolution equations for an RTP are translationally invariant, Eqs.~\eqref{ptot_sym}~and~\eqref{lambda} directly yield
\begin{eqnarray}
\tilde G^0(x,-z,s)&=& \frac{e^{-\frac{\left|
   x+z\right|  \sqrt{s (s+2
   \gamma )}}{v}} \sqrt{s (s+2
   \gamma )}}{2 v s},\quad z=-x_i,
\label{greens_functions_laplace}
\end{eqnarray}
where $G^0(x,-z,s)$ is the Laplace transform of the Green's function for an RTP initialized symmetrically in $+$ and $-$ states. It is useful to compute certain quantities related to the Green's function for the analytical calculations presented in the main text.
Let us first define
\begin{eqnarray}
U^0(z,t) = \int_0^\infty dx~G^0(x,-z,t), \quad z\geq 0 ,
\label{uzt}
\end{eqnarray}
as the integral of the Green's function $G^0(x,-z,t)$ over the half-infinite line. 
Using the expression in Eq.~\eqref{greens_functions_laplace}, we obtain the exact expression for $U^0(z,t)$ in Laplace space as
\begin{eqnarray}
\tilde U^0(z,s) &=& \frac{\exp\left(-z\frac{\sqrt{s(s+2\gamma)}}{v} \right)}{2s}.
\label{uzs}
\end{eqnarray}
The integral of the function $\tilde U^0(z,s)$ over $z$ can be easily computed as
\begin{eqnarray}
\int_0^\infty dz~\tilde U^0(z,s) =\frac{v}{2 s^{3/2}\sqrt{(s+2 \gamma)}}. 
\label{Us_int}
\end{eqnarray}
We next define 
\begin{equation}
\tilde V^0(z,s)=\mathcal{L}\left[{U^0(z,t)}^2\right],
\label{Vs_def}
\end{equation}
as the Laplace transform of the square of the function $U^0(z,t)$. The integral of the function $\tilde V^0(z,s)$ is another useful quantity that enters into the computations presented in the main text. It can be shown that 
\begin{eqnarray}
\int_0^\infty dz~\tilde V^0(z,s) =\frac{ v}{2 s^{3/2}} \left(\frac{1}{\sqrt{s+2 \gamma }}-\frac{\sqrt{s+4 \gamma }}{2( s+2\gamma)}\right). 
\label{Vs_int}
\end{eqnarray}

\subsection{Asymmetric initial conditions}

We next compute the Green's functions for initial conditions where the particle is initialized in either of the two states $+$ or $-$. We analyze these cases separately.
\subsubsection{Particle initialized in $+$ state:}
We first consider asymmetric initial conditions of the form
\begin{equation}
P_+(x,0) = \delta(x),\quad P_-(x,0) = 0.
\label{asym_cond_plus}
\end{equation}
Here, the particle starts from $+$ velocity state at time $t=0$. In Fourier space, these initial conditions assume the form $\ket{\tilde P_m (k,t=0)}~=~{\begin{pmatrix}
1~
0
\end{pmatrix}}^T$.
Solving Eq.~(\ref{matrix11a}) along with these initial conditions yields the exact expression for the Fourier-Laplace transform of the total probability density as
\begin{equation}
\tilde P (k,s)=\frac{s+2 \gamma+i vk }{s (s+2 \gamma )+v^2 k^2},
\label{fl_plus}
\end{equation}
where  $\tilde P(k,s)~=~\tilde P_+(k,s)+\tilde P_-(k,s)$. We next invert the Fourier transform in the above equation and this yields
\begin{equation}
\tilde P(x,s) =
=  \frac{{\rm e}^{- |x|\lambda}}{2s}\left(\lambda+\text{sgn}(x)\frac{s}{v}\right),
\label{ptot_asym_plus}
\end{equation}
where \text{sgn} denote the sign function and the expression for $\lambda$ is provided in Eq.~\eqref{lambda}.

\subsubsection{Particle initialized in $-$ state}
We next consider asymmetric initial conditions of the form
\begin{equation}
P_-(x,0) = \delta(x),\quad P_+(x,0) = 0.
\label{asym_cond_minus}
\end{equation}
Here, the particle starts from $-$ velocity state at time $t=0$. In Fourier space, we have $\ket{\tilde P_m (k,t=0)}~=~{\begin{pmatrix}
0~
1
\end{pmatrix}}^T$.
Solving Eq.~(\ref{matrix11a}) along with these initial conditions yields the exact expression for the Fourier-Laplace transform of the total probability density as
\begin{equation}
\tilde P (k,s)=\frac{s+2 \gamma-i vk }{s (s+2 \gamma )+v^2 k^2},
\label{fl_minus}
\end{equation}
where  $\tilde P(k,s)~=~\tilde P_+(k,s)+\tilde P_-(k,s)$. After inverting the Fourier transform in the above equation, we directly obtain
\begin{equation}
\tilde P(x,s) =   \frac{{\rm e}^{- |x|\lambda}}{2s}\left(\lambda-\text{sgn}(x)\frac{s}{v}\right),
\label{ptot_asym_minus}
\end{equation}
where \text{sgn} denote the sign function and the expression for $\lambda$ is provided in Eq.~\eqref{lambda}.\\

Let us define the Green's function $G^\pm(x,x_i,t)$ as the probability density of finding an RTP at location $x$ at time $t$, starting from the location $x_i$ at time $t = 0$ in the velocity state $\pm$. Here, the superscript $ ``\pm"$ denotes the asymmetric case where the particle starts from the $+$ or $-$ state at time $t=0$ with probability $1$. Since the evolution equations for an RTP are translationally invariant, Eqs.~\eqref{ptot_asym_plus},~\eqref{ptot_asym_minus}~and~\eqref{lambda} directly yield
\begin{eqnarray}
\tilde G^\pm(x,-z,s) &=& \frac{e^{-\frac{\left|
   x+z\right|  \sqrt{s (s+2
   \gamma )}}{v}}
   \left(\sqrt{s (s+2 \gamma
   )} \pm s~
   \text{sgn}(x+z)\right)}{2 v
   s},\quad z=-x_i.
\label{greens_functions_laplace_plus_minus}
\end{eqnarray}
Let us define
\begin{eqnarray}
U^\pm(z,t) = \int_0^\infty dx~ G^\pm(x,-z,t), \quad z\geq 0 ,
\label{uzt_plus_minus}
\end{eqnarray}
as the integral of the Green's function $G^\pm(x,-z,t)$ over the half-infinite line.
Using the expression in Eq.~\eqref{greens_functions_laplace_plus_minus}, we obtain the exact expression for the Laplace transform of $U^\pm(z,t)$ as
\begin{eqnarray}
\tilde U^\pm(z,s) &=& \frac{e^{-\frac{z \sqrt{s (s+2
   \gamma )}}{v}}}{2
   s}
   \left(1 \pm \frac{s}{\sqrt{s
   (s+2 \gamma )}}\right).
\label{uzs_plus_minus}
\end{eqnarray}
The integral of the function $\tilde U^\pm(z,s)$ over $z$ yields
\begin{eqnarray}
\int_0^\infty dz~\tilde U^\pm(z,s) =\frac{ v \left(\sqrt{s (s+2
   \gamma )} \pm s\right)}{2 s^2
   (s+2 \gamma )} . 
   \label{Uplus_minus_int}
\end{eqnarray}
We next define 
\begin{equation}
\tilde V^\pm(z,s)=\mathcal{L}\left[{U^\pm(z,t)}^2\right],
\label{Vplus_minus_def}
\end{equation}
as the Laplace transform of the square of the function $U^\pm(z,t)$. It can be shown that 
\begin{eqnarray}
\hspace{-2.5 cm}
\int_0^\infty dz~\tilde V^\pm(z,s) =\frac{v }{s (s+2 \gamma )}\left(\pm \frac{1}{4}+\frac{1}{2} \sqrt{\frac{s+2 \gamma }{s}}-\frac{\gamma }{\sqrt{s
   (s+4 \gamma )}} \pm \frac{K\left(-\frac{8 \gamma  (s+2 \gamma )}{s^2}\right)}{2 \pi
   }\right). 
   \label{Vplus_minus_int}
\end{eqnarray}
We also define 
\begin{equation}
\tilde V^{\text{cross}}(z,s)=\mathcal{L}\left[U^+(z,t)U^-(z,t)\right],
\end{equation}
as the Laplace transform of the product of the functions $U^+(z,t)$ and $U^-(z,t)$.
Using Eq.~\eqref{U_0_def} in the main text and the definitions provided in Eqs.~\eqref{Vs_def} and \eqref{Vplus_minus_def} we obtain
\begin{equation}
\tilde V^{\text{cross}}(z,s)=2 \tilde V^{0}(z,s)-\frac{1}{2}\left(\tilde V^{+}(z,s)+\tilde V^{-}(z,s) \right).
\end{equation}
Integrating the above equation over $z$ and substituting the expressions provided in Eqs.~\eqref{Vs_int} and \eqref{Vplus_minus_int} yield
\begin{equation}
\hspace{-0.6 cm}
\int_0^\infty dz~\tilde V^{\text{cross}}(z,s) =\frac{v }{\sqrt{s} (s+2 \gamma )^{3/2}}\left(\frac{1}{2}+\frac{\gamma }{s}+\frac{\gamma  \sqrt{\frac{s (s+2 \gamma )}{s+4
   \gamma }}}{s^{3/2}}-\frac{\sqrt{s (s+2 \gamma ) (s+4 \gamma )}}{2
   s^{3/2}}\right).
    \label{Vcross_int}
\end{equation}
\section{Derivation of the expressions of the variance for different initial conditions}

We next turn to the computation of the exact expression for $\mu(t)=\sigma_{a,a}^2=\sigma_{a,q}^2$ defined in Eq.~\eqref{mu_t} of the main text. This is equal to the mean of $Q$ for all initial conditions and the variance for annealed density and annealed or quenched magnetization initial conditions. In Laplace space, the expression in Eq.~\eqref{mu_t} of the main text translates to

\begin{equation}
  \tilde \mu(s)=\rho \int_0^{\infty} dz ~  \tilde U^{0}(z,s).
  \label{mu_s}
\end{equation}
Substituting Eq.~\eqref{Us_int} in the above equation and inverting the Laplace transform yields the exact expression for the variance for annealed density and annealed or quenched magnetization initial conditions. This expression is provided in Eq.~\eqref{var_a_density} of the main text. 

To compute the variance $\sigma_{q,q}^2$ for quenched density and quenched magnetization initial conditions, we rewrite the expression provided in Eq.~\eqref{var_q_q_def} of the main text in Laplace space as
\begin{eqnarray}
\tilde \sigma_{q,q}^2 (s) =\frac{\rho}{2} \left( \tilde T_1 (s)+\tilde T_2 (s) \right),
\label{var_q_q_s}
\end{eqnarray}
where 
\begin{equation}
 \tilde T_1 (s)   = \mathcal{L}\left[ \int_0^\infty dz~\left[ U^+(z,t)(1-U^+(z,t)) \right]  \right]=\int_0^\infty dz~\left[\tilde U^+(z,s)-\tilde V^+(z,s)\right],
\end{equation}
and
\begin{equation}
 \tilde T_2 (s)   = \mathcal{L}\left[ \int_0^\infty dz~\left[U^-(z,t)(1-U^-(z,t)) \right]   \right]=\int_0^\infty dz~\left[\tilde U^-(z,s)-\tilde V^-(z,s)\right],
\end{equation}
where $\tilde V^+(z,s)$ and $\tilde V^-(z,s)$ are defined in Eq.~\eqref{Vplus_minus_def}.
Substituting the expressions provided in equations~\eqref{Uplus_minus_int}~and~\eqref{Vplus_minus_int} in the above equations, we obtain
\begin{eqnarray}
\tilde T_1 (s)  = 
\frac{v}{4 s (s+2 \gamma )}+\frac{v \gamma }{s (s+2 \gamma ) \sqrt{s (s+4 \gamma )}}-\frac{v
   K\left(-\frac{8 \gamma  (s+2 \gamma )}{s^2}\right)}{2 \pi  s (s+2 \gamma )},
   \label{T1}
\end{eqnarray}
and
\begin{eqnarray}
\tilde T_2(s)  = 
-\frac{v}{4 s (s+2 \gamma )}+\frac{v \gamma }{s (s+2 \gamma ) \sqrt{s (s+4 \gamma )}}+\frac{v
   K\left(-\frac{8 \gamma  (s+2 \gamma )}{s^2}\right)}{2 \pi  s (s+2 \gamma )}.
   \label{T2}
\end{eqnarray}
Combining equations~\eqref{var_q_q_s},~\eqref{T1}~and~\eqref{T2}~and~inverting the resultant Laplace transform yields the result provided in Eq.~\eqref{var_q_q} of the main text. 

We next compute the explicit expression of the variance $\sigma_{q,a}^2$ for quenched density and annealed magnetization initial conditions. Using the definitions provided in Eqs.~\eqref{U_0_def}~and~\eqref{var_q_a_def} of the main text, we obtain the expression for the variance in Laplace space as
\begin{eqnarray}
\tilde \sigma_{q,a}^2 (s) =\frac{\rho}{2} \int_0^\infty dz~\Big[\tilde U^{+}(z,s)+\tilde U^{-}(z,s)-\frac{1}{2} \tilde V^{+}(z,s)
- \frac{1}{2}\tilde V^{-}(z,s)-\tilde V^{\text{cross}}(z,s)\Big] .
\label{var_q_a_laplace_exp}
\end{eqnarray}
Each term in the above expression can be computed explicitly. Combining results from equations~\eqref{Uplus_minus_int},~\eqref{Vplus_minus_int}~and~\eqref{Vcross_int}, we directly obtain the expression for the variance in Laplace space inverting which yields the result in Eq.~\eqref{var_q_a} of the main text.

\clearpage
\end{widetext}

\end{document}